\documentstyle[12pt,aasms4]{article}
\slugcomment{Second Revised Version Submitted to PASP the June 16, 1998}

\lefthead{Musazzi et al.}
\righthead{V798 Cyg and V831 Tau}

\begin{document}
\def\fu{$f_1~$}
\def\t{$\pm$}
\def\fr{$f_1/f_2~$}
\def\fd{$f_2~$}
\def\fdu{$f_2 - 2f_1~$}
\def\fp{$f_1 + f_2~$}
\def\fm{$f_2 - f_1~$}
\def\cd{cd$^{-1}$}
\def\cds{cd$^{-1}$\,}
\def\kms{km~s$^{-1}$}
\def\kmss{km~s$^{-1}$\,}
\def\I{\'\i}
\def\salp{\vskip 0.3truecm}
\title{CCD Photometry of the High Amplitude $\delta$ Scuti Stars\\
 V798 Cyg and V831 Tau}
\author{F. Musazzi\altaffilmark{1}, E. Poretti\altaffilmark{1} and 
S. Covino\altaffilmark{1}}
\affil{Osservatorio Astronomico di Brera, Merate, Italy\\
poretti/covino@merate.mi.astro.it}
\and
\author{A. Arellano Ferro\altaffilmark{2}}
\affil{Departamento de Astronom\I a, Universidad de Guanajuato, M\'exico\\
armando@astro.ugto.mx}
\altaffiltext{1}{Osservatorio Astronomico di Brera, Via Bianchi 46,
I-23807 Merate, Italy}
\altaffiltext{2}{Departamento de Astronom\I a, Universidad de Guanajuato, 
Apdo Postal 144, Guanajuato, Gto. M\'exico 36000}

\begin{abstract} New CCD measurements on the two high--amplitude $\delta$
Sct stars V798 Cyg and V831 Tau were carried out. The double--mode pulsation
of V798 Cyg is demonstrated beyond any doubt. The \fr ratio of
0.800 is confirmed to be related to the unusual shape of the \fu light
curve. The properties of the Fourier parameters were revisited also by
considering the new light curve of V831 Tau. In particular, the classical
double--mode pulsators cannot fill the gap in the $R_{21}$ distribution.
Two new variable stars were discovered in the field of V798 Cyg.
\end{abstract}

% The different journals have different requirements for keywords.  The
% keywords.apj file, found on aas.org in the pubs/aastex-misc directory, 
% contains a list of keywords used with the ApJ and Letters.  These are 
% usually assigned by the editor, but authors may include them in their 
% manuscripts if they wish. 

\section{Introduction}
In the recent past an increasing number of theoretical and observational
studies were devoted to the $\delta$ Sct stars showing light curves having
a $V$ amplitude larger than $\sim$0.20 mag and pulsating in one or two
stable frequencies (HADS, see also Petersen \& Christensen--Daalsgard
1996). They were mainly concerned with the
understanding of their pulsational and physical
properties. Rodr\I guez et al. (1996) investigated the nature of the large
amplitude modes, concluding that they are radial; Garrido \& Rodr\I guez (1996)
refined this conclusion, finding possible nonradial terms having very small
amplitudes
in SX Phe and DY Peg. Breger \& Pamyatnykh (1998) 
found evidence of period changes in a number of HADS, but
it was not possible to relate them with stellar evolution; hence, their nature
is still unclear.

Antonello et al. (1986) and Poretti et al. (1990) proposed
a synthetic description of the HADS light curves, trying to obtain
their quantitative classification. The result was successfull, since regular
trends were observed in the Fourier parameter space, as in the $\phi_{21}-P$ and
$\phi_{31}-P$ planes. A bimodal distribution was observed in $R_{21}-P$ plots
and no satisfactory explanation can be given at the moment, since the action of
a resonance is not confirmed by the phase plots. Moreover, Poretti \& Antonello
(1988) emphasized the similarities between the light curves of V1719 Cyg,
V798 Cyg and V974 Oph which show a descending branch steeper than the
ascending one.  Later, V974 Oph was dropped out from this short list since
it is actually a multiperiodic variable (Poretti \& Mantegazza 1992),

Recently, Hintz \& Joner (1997) reported new results on CY Aqr, XX Cyg and 
V798 Cyg. Their result for the first two stars confirm  the Fourier parameters
obtained by Antonello et al. (1986). On the other hand, 
Hintz \& Joner questioned the double--mode nature of V798 Cyg 
suggested by Poretti \& Antonello (1988). In this paper
we present new CCD data which corroborate the double--mode hypothesis;
we also rediscuss the properties of the Fourier parameters considering the first
CCD measurements of the short period variable V831 Tau.

\section{New observations}
After the detection of a possible second periodicity in the light curve of
V798 Cyg, we planned new photometric measurements to confirm it. 135 CCD 
images were obtained with the 152--cm Cassini telescope 
of the Bologna Astronomical Observatory, located at Loiano, on four nights
between 1987 August 29 and 1987 September 4.  The observations were obtained
through a  $V$ filter and the exposure time was 20 sec (15 sec for the first
20 images).
Figure 1 shows a typical frame.
%\placefigure{fig1}

More recently,  we carried out an observing run on some extragalactic objects
with the 1.5 m telescope at San Pedro M\'artir Observatory (Baja California, Mexico) in December 1996.
When seeing
conditions were not adequate (owing to strong wind) to take CCD frames on
20th--mag objects,
we considered the 17th--mag variable star V831 Tau (named as GR 269 and ZB 33
in Romano 1973 and Huang 1982, respectively) as a good backup
programme. Exposure times ranged from 7 to 15 minutes depending on weather
conditions. 81 CCD $V$ images were obtained on six nights from December 11
to 18, 1996. The rapid and large
amplitude variability made V831 Tau  an interesting object since the light
curve reported by Huang (1982) is sinusoidal and such a shape
is quite unexpected for short period variables. Figure 2 shows a typical
CCD frame of the field.
%\placefigure{fig2}
The data reduction was performed by using the {\sc midas}
package; both photometry aperture and {\sc daophot} routines were used. Because
the target stars in our frames are very bright, the methods are equivalent.
For a full description of the analysis, see Musazzi (1997).
The results described here are obtained using a photometric  aperture of
14 arcsec for the San Pedro M\'artir frames and 10 arcsec for the Loiano frames.
Photometry was performed for all the stars in each field; differential
magnitudes were calculated with respect to the adequate comparison
star. In the case of V798 Cyg (Fig. 1), $\Delta V$ values were
derived with respect to star 2; they were 
tranformed into an instrumental $V$ system by assuming $<V>$=12.504 for
V798 Cyg (Poretti \& Antonello 1988). 
An estimation of the errors was achieved from the standard deviation of the
measurements of star 4
in the V798 Cyg frames and of the star 3 ones in the V831
Tau frames: the whole datasets yielded 5 and 7 mmag, respectively.

As a by--product of our analysis, we detected two new variables in the rich 
field of V798 Cyg. The first one, labelled ``6" in Fig.~1, is probably an
eclipsing binary.  Figure~3
shows its light curve on the three longest nights: the regular decrease
in the top panel is about 0.5 mag, while on the other nights it is constant
at maximum level. The second new variable, labelled ``8" in Fig.~1,
is probably a multiperiodic pulsating star. Figure~4 shows its light
curve: cycle--to--cycle variations
are clearly evident. A frequency analysis revealed two possible terms at
13.95 and 19.58 \cd, but further observations are necessary to disclose its
true nature.
%\placefigure{fig3}
%\placefigure{fig4}
\section{Frequency analysis}
The least--squares power spectrum method (Vani\^cek 1971) was used to detect
the frequency content of the light curves. This method allows each constituent
to be detected one by one: in the search for a new term, only the frequencies
of the previously identified terms (the known constituents, k.c's) are taken
into account, since their amplitudes and phases are recalculated as
unknowns in the least--squares routine.
We routinely apply this algorithm to multiperiodic
$\delta$ Sct stars; recently, Pardo \& Poretti (1997) used it to detect the
frequency content of the light curves of double--mode Cepheids.

\subsection{V798 Cyg}
The new CCD measurements were analysed to confirm the presence of a second
excited mode. The frequency analysis immediately evidenced the peak at 5.13
\cds (Fig.5, top panel), corresponding to the well--known first period (Poretti
\& Antonello 1988). There is no doubt on it, since when introducing the
neighbouring 
alias (4.13 or 6.13 \cd) residual signal is left at 5.13 \cd.
By proceeding in the analysis the
harmonics 2$f_1$, 3\fu and 4\fu (or its aliases) were detected as the highest
peaks. This is not
surprising since the light curve on \fu is known to be quite asymmetrical.

The crucial point was to verify what happened when \fu and its first three
harmonics were introduced as k.c's in the search for new frequencies. The
middle panel of Fig.~5 shows the power spectrum we obtained: the structure
centered around 6.41 \cds is clearly visible. This panel can be compared
with the Fig. 5 in Poretti \& Antonello (1988): in that figure the uncertainty
between 6.41 and  7.41 \cds is strong due to the interaction between
noise effects and spectral window, since in all these cases the time
sampling is not
particularly good. We can conclude that the spectral analysis of the new,
more accurate, CCD data suggests 6.41 \cds as the true value for the \fd term.
Figure 6
shows the light curves obtained by subtracting from the data the curve related
to one of the two frequencies and then plotted against the phase related to
the other one. As can be seen the CCD light curves in Fig. 6 are of better
 quality with respect to the photolelectric ones (Fig. 6 in Poretti \&
Antonello 1988); in particular the scatter around the minimum of the light
curve on \fu is considerably reduced and it can be ascribed to the low
signal in the photoelectric measurements when the star is very faint.
Also considering the middle panel of Fig. 5,
no doubt is left on the fact that a second periodicity  does exist. The two
frequency analyses carried out on two independent datasets show the same
structure over the whole spectral pattern (\fu and its three harmonics, $f_2$),
strongly supporting its real nature.

It is now necessary to understand
why Hintz \& Joner (1997) failed to find the \fd term. It
should be noted that they did not report any power spectrum, 
did not give any detail about the method used to analyze their time series
and no observing log is presented. A careful rediscussion of their
measurements was therefore considered necessary.
 The authors kindly put at
our disposal their unpublished measurements and we processed them. 
The data were obtained on only two nights spaced by one night.
As a result, the peaks in the power spectrum are quite large and the aliases
at 0.5 \cds are also prominent. Since  Hintz \& Joner (1997) 
performed a prewhitening at each step,  some power belonging to \fd may
be arbitrarily subtracted when removing the \fu light curve,
owing to the leakage between the large and
numerous peaks. Hence, the importance of \fd has been underevaluated.
To verify this hypothesis, we analyzed the Hintz \& Joner
measurements by using our method, which does not require any
prewhitening: the bottom panel of Fig.~5 shows the power spectrum we obtained.
In spite of its rather poor look 
(owing to the bad spectral window), the presence of a
signal around 5--7 \cds is clearly evidenced. We can conclude that 
their measurements also support the presence of a second periodicity.
%\placefigure{fig5}
%\placefigure{fig6}
Why did they missed it? They observed on two nights only and the
influence of \fd shifted the light curve built on \fu=5.1343 \cd. Hence
they could obtain a good fit  with a slightly
different value for the frequency (5.1261 \cd); the prewhitening of
\fu in their data may have subtracted power from $f_2$.
However, the 5.1261 \cds  value is fully inadequate to fit all the available data
on V798 Cyg (not only our CCD and photoelectric sets, but also the former
photographic measurements); this different frequency value can be considered
as an artifact of the analysis, just good enough to link the light curves of
two nights only of a double--mode
variable with a single (but wrong) period. To verify once more the action of
the \fd term in the Hintz \& Joner data, Fig.~7 shows the light curves of
\fu and \fd as obtained from the
Hintz \& Joner measurements: the sine--shaped \fd light
curve is quite evident. When comparing Figs. 6 and 7, the light curves on
the \fu term seem to be different in shape at maximum light, but this must be
considered with caution because our  CCD measurements cover only
one cycle around the maximum. Indeed, the Fourier parameters are very similar.
They are summarized in Table 1, together with the parameters of the
least--squares  fit. The small differences in the amplitudes can be entirely
due to the different instrumental systems; this fact hampered us from
merging all the measurements in a reliable way.
%\placetable{tbl-1}

%\placefigure{fig7}
\subsection{V831 Tau}
The analysis of the 81  CCD $V$ measurements of V831 Tau was much easier
than in the previous case. The terms  \fu=15.551 \cd, 2$f_1$, 3$f_1$, 4$f_1~$ were
clearly identified in the power spectrum; they correspond to the already 
known period of 0.0643 d. The fit with these four terms
leaves a rms residual of 0.014 mag (Tab. 2); it is larger than the expected one,
but we have to consider that V831 Tau is 2 mag fainter than star 2 in 
Fig. 2. Moreover, owing to the large amplitude and short
period,  during the exposure time V831 Tau showed an
intrinsic, not linear variability and hence ascribing the observed flux to the
middle time is not always an exact procedure.
The residual power spectrum does not allow any identification of a second
period or a higher harmonic.
Figure 8 shows the light curve; as can be noted, it is quite asymmetrical
and the ascending branch is very steep. Moreover,  it is quite regular; a 
small scatter is observed only at the turning points, where the shape of
the light curve is changing non--linearly during the exposures. The sine--shaped
light curve presented by Huang (1982) is not confirmed; it probably
originated from the very long photographic exposures. It should be noted that
its period is one of the shortest among HADS, very close to that of CY Aqr.
%\placetable{tab-2}
%\placefigure{fig8}
\section{Discussion}
\subsection{The search for a second term}
From a methodological point of view, it was once more demonstrated that
the detection of a second periodicity is not an easy task when it has a
very small amplitude. A careful frequency analysis must be performed and
the quality of the fit cannot be considered a powerful diagnostic of the
absence of a second term. Harmonics and cross--coupling terms can be
detected one by one using our iterative least--squares technique, as
Pardo \& Poretti (1997) extensively demonstrated in the case of double--mode
Cepheids. In such a context, the approach of Hintz \& Joner (1997) aiming
at searching for the best solution is inadequate and is not recommended
for future works: as it was demonstrated by the power spectra and the light
curves (Figs. 3, 4 and 5), the second term \fd is clearly evident
in the measurements of V798 Cyg. Moreover, it should be emphasized that 
Hintz \& Joner
did not verify if their solution could fit the previous datasets. In such
a case they could easily verify that their period was  wrong; from a general
point of view, two short nights cannot invalidate the well--proven results
obtained on a larger baseline.

When the presence of a second periodicity is claimed or refused on the basis
of qualitative statements, some contradictions may arise. As an example, Hintz
\& Joner (1997) considered V567 Oph as a possible double--mode pulsator because
Powell et al. (1990) had found amplitude variations and phase shifts.
Since Powell et al. (1990) did not detect any second term in
the power spectrum of V567 Oph, we cannot consider it as a double--mode 
pulsator; moreover, Poretti et al. (1990) found its light curve to be
very stable. On the
other hand, they considered CY Aqr as a monoperiodic pulsator, but this star
shows the same cycle--to--cycle variations (McNamara et al., 1996, and 
references therein).
Such variations are the definition of the well--known (but also not 
well--understood) Blazhko effect, typical of HADS
and RR Lyr stars. As a rule, it is not possible to explain it as the action
of a second periodicity, since frequency analysis of well--observed stars
failed to detect it.  As a last
comment, it should be noted that Hintz \& Joner (1997) considered V1719 Cyg as a
suspected double--mode pulsator in their discussion; however, 
as clearly demonstrated by Poretti \&
Antonello (1988), the double--mode pulsation is quite evident and there
is no reason to doubt it.
In our opinion, only a frequency
power spectrum obtained on the basis of good quality data (both for precision
and for time sampling) should be
used to argue in favour of or against a double--mode behaviour.

\section{The Fourier decomposition}
To perform the decomposition,
we fitted the original measurements by the formula
\begin{equation}
\Delta V(t)= \Delta V_o + \sum_i^n {A_i \cos [2\pi if_1  (t-T_o) +\phi_i ]}+\\
\sum_i^m {A_i \cos [2\pi if_2  (t-T_o) +\phi_i ]}
\end{equation}
In the case of V831 Tau, \fu, 2$f_1$, 3$f_1$, 4\fu were identified in the 
frequency analysis, i.e. there was no\fd term.
The corresponding phase  $\phi_{ij}=j\phi_i-i\phi_j$ and amplitude
$R_{ij}=A_i/A_j$ parameters can be derived. In the case of V798 Cyg, the
$f_1$, 2$f_1$, 3$f_1$, 4\fu and \fd
terms were identified; because $m$=1, no Fourier parameter
can be calculated for the \fd  term.
\subsection{V831 Tau}
The Fourier decomposition of the $V$ CCD measurements yields
$\phi_{21}=\phi_2-2\phi_1$=3.75\t0.09 rad and $R_{21}=A_2/A_1$=0.36\t0.03
for V831 Tau. This star provides a good confirmation of the regular trend
in the short period region of the $\phi_{21}-P$ plots (see Fig. 8 in
Poretti et al. 1990): in fact, from 0.10
to 0.04 d we can see a drop of the $\phi_{21}$ values from 4.0 rad to
about 3.0 rad. The reason for such an abrupt change in the light curve shape
is unknown at present. V831 Tau takes place in the group having
large $R_{21}$ values, as is often the case for short period variables. The
asymmetrical light curve behaviour is the expected one for such a variable
and is more likely than the previously reported sinusoidal shape. 
\subsection{V798 Cyg and the double--mode pulsators}
The least--squares fits reported in Tab. 1 allowed us to obtain 
$\phi_{21}$  values in good agreement with each other: 2.53\t0.08,
2.64\t0.05, 2.70\t0.06 rad, respectively. Poretti \& Antonello (1988) determined
a value of 2.52\t0.05 rad from $uvby$ photometry of 
 V1719 Cyg. Since the two $R_{21}$ values are also very similar (0.18,
0.19, 0.17 in the three datasets of V798 Cyg; 0.20 for V1719
Cyg), the close similarity between their \fu light curves is now 
a well--established result. The \fd light curves are slightly different;
in  V798 Cyg the \fd  amplitude is shallow and hence cross--coupling terms are 
not detected, while they are observed in the V1719 Cyg \fd light curve.
The \fu/\fd ratio is 0.7998\t0.0001 for V1719 Cyg and 0.8012\t0.0002 for V798
Cyg: since the separation is about one order of magnitude greater than the
formal errors, a slight dependence from \fu is suggested.

Hintz \& Joner (1997) argued that double--mode pulsators can fill the gap
in the $R_{21}$ distribution, supplying values around 0.27. They quoted the
Rodr\I guez et al. (1992) results to support their thesis and in particular the
0.27 value obtained in the cases of BP Peg and RV Ari. However, as can be 
easily verified, this intriguing hypothesis is wrong as Rodr\I guez et al.
report 0.303 for RV Ari and 0.315 for BP Peg; the same values are obtained
from their least--squares fit. So the gap around 0.27 still exists. 

It should be emphasized once more that the \fr ratio for AE UMa, BP
Peg and RV Ari is about 0.773. 
Petersen \& Christensen--Daalsgard (1996) showed how the 
0.77 ratio is in good agreement with the predictions of
the theoretical models for the fundamental
($\Pi_0$) and the first overtone ($\Pi_1$) radial modes. Small deviations
can be obtained varying the metallicity
or assuming different $M-L$ relationships, but for $\log \Pi_0 < -0.8$ 
all the theoretical values are within 0.765 and 0.790. When discussing the case
of VZ Cnc, Petersen \& H\o g (1998) argued against
the hypothesis of a $\Pi_1/\Pi_0$ ratio modified by secondary parameters as
diffusion or rotation and considered the 0.800 ratio as an indicator of the
pulsation in a different couple of modes, i.e. the second ($\Pi_2$) 
and the first radial overtone. In this scenario, we should consider
V1719 Cyg and V798 Cyg as two $\Pi_2/\Pi_1$ pulsators; however, we 
pointed out the unusual shape of their \fu light curve (considering it both
a $\Pi_0$ or a $\Pi_1$ mode), very different from
that of other double--mode pulsating stars, both HADS and Cepheids.
Since in the case of Cepheids the deviations from regular trends in the Fourier
parameters planes were recognized as indicators of different physical
conditions (other pulsation modes, resonances; Poretti 1994 and references
therein, Welch et al. 1995), further theoretical investigations are needed to
give a full description of the observed peculiarities emphasized above.

\section{Conclusions}
The possibility to perform the frequency analysis of the  light variation
of V798 Cyg on the basis of a more accurate set of data and the Fourier
decomposition of the new light curve of V831 Tau allowed us to
improve the knowledge of the HADS. The following points are now better defined:
\begin{enumerate}

\item  V798 Cyg 
is confirmed to be a double--mode pulsator and its similarity with V1719 Cyg
definitely established: the peculiar light curve on \fu (descending branch
steeper than the observed one) and the double--mode pulsation with ratio 0.800
have to be considered together in modelling these stars.
\item The detection of additional periodicities in light curves is confirmed
to be a delicate matter. A thorough discussion would be needed to invalidate
previous results estalished on the basis of better time--series, as
otherwise confusing results are obtained, as in the case of V798 Cyg.
\item The short--period
variable V831 Tau shows a light curve very similar to other stars having the
same period; the previously reported sinusoidal light curve is not
confirmed. The light curves of HADS towards short periods have a very regular
progression in the Fourier parameters space.
\item The gap around 0.25--0.30 in the $R_{21}$ distribution of HADS
light curves is confirmed. Hintz \& Joner (1997)
claimed that it can be filled by the values obtained from double--mode
pulsators, but the available photometry does not
support that claim which seems the result from improper handling of the data.
\end{enumerate}

\acknowledgments
E.~Poretti wish to thank V. Zitelli for the help
in the observing run in Loiano. E.~Poretti and A.~Arellano Ferro acknowledge
the financial support of a  CNR--CONACyT contract.
The authors wish to thank J. Vialle for the improvement of the English form
of the manuscript.

\clearpage
\begin{figure}
\epsscale{0.5}
\plotone{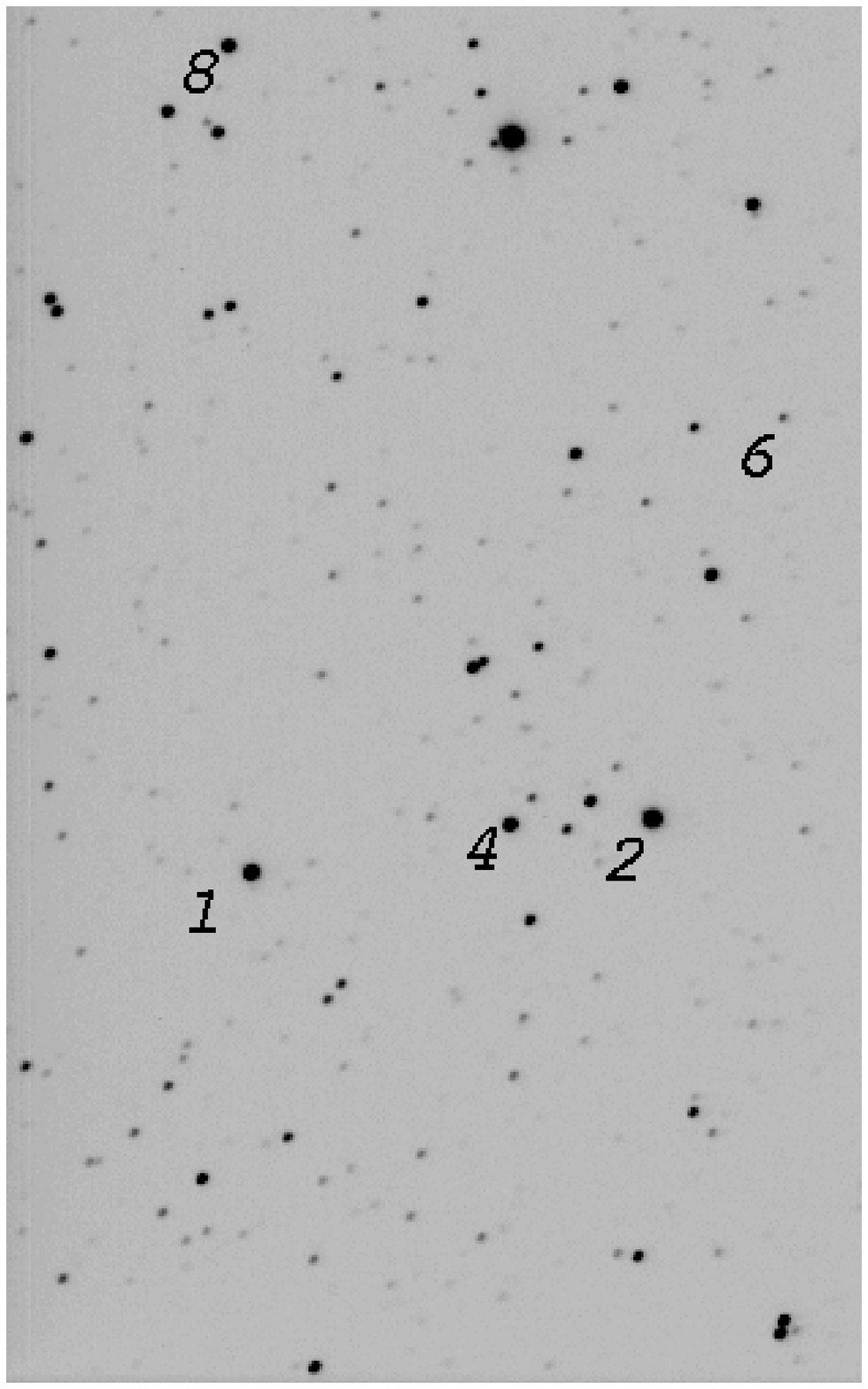}
\caption{The typical frame of the V798 Cyg field. North is down and west is
right. The field is 2.5~x~4.4 arcmin$^2$. V798 Cyg is labelled as star 1; stars
2 and 4 are the selected comparison stars; star 6 is probably a new eclipsing
binary; star 8 is probably  a multiperiodic new variable star. \label{fig1}}
\end{figure}

\clearpage
\begin{figure}
\epsscale{0.5}
\plotone{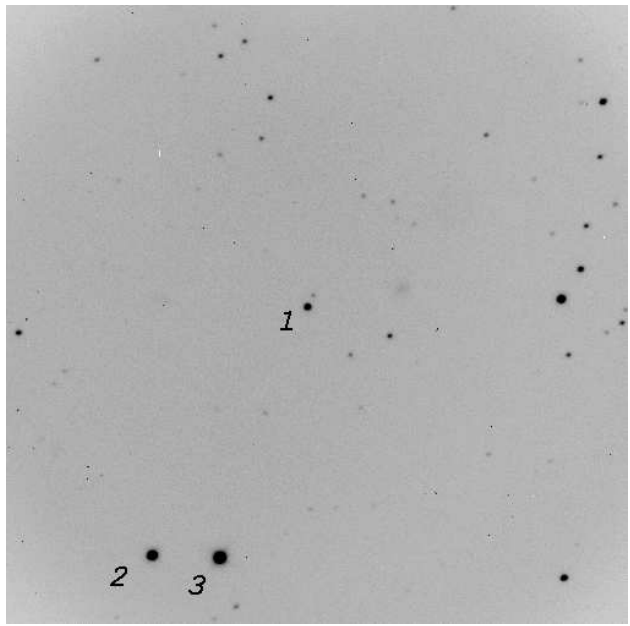}
\caption{The typical frame of the V831 Tau field. North is up and west is
left. The field is about 7~x~7 arcmin$^2$. V831 Tau is labelled as star 1; star
3, the brightest of the field, is the comparison star. \label{fig2}}
\end{figure}

\clearpage
\begin{figure}
\epsscale{0.58}
\plotone{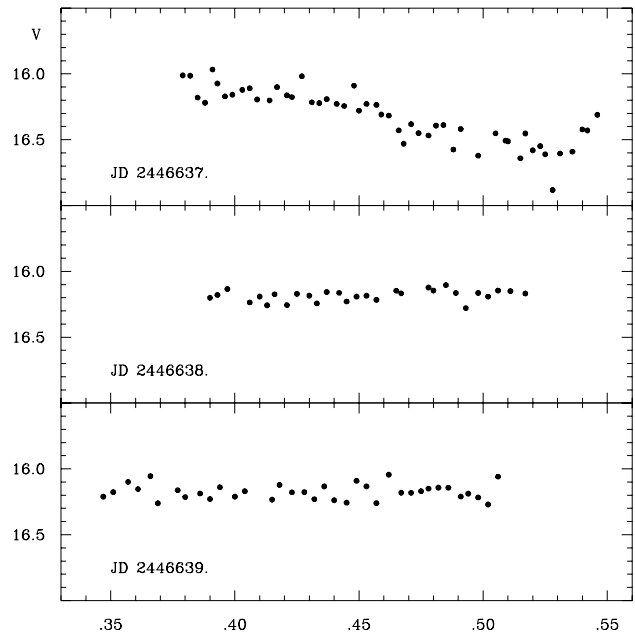}
\caption{Light curves of a probable new eclipsing variable in the field of
V798 Cyg (star 6 in Fig. 1). \label{fig3}}
\end{figure}

\clearpage
\begin{figure}
\epsscale{0.58}
\plotone{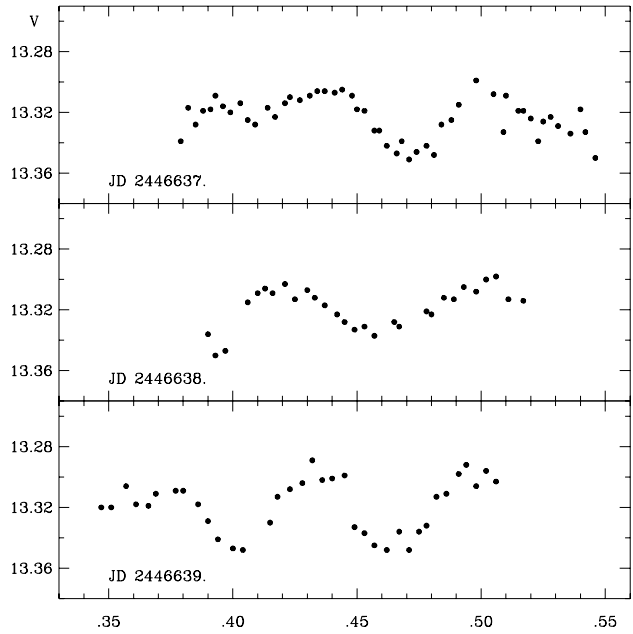}
\caption{Light curves of a possible  new multiperiodic variable in the
field of V798 Cyg (star 8 in Fig. 1). \label{fig4}}
\end{figure}

\clearpage

\begin{figure}
\epsscale{0.56}
\plotone{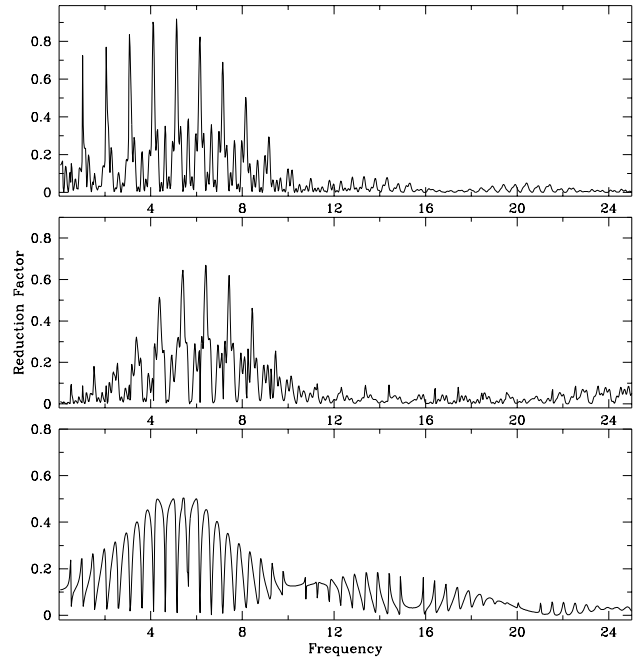}
\caption{Power spectra of V798 Cyg. Top: Loiano CCD measurements,
no k.c.'s. Middle: Loiano CCD measurements, \fu, 2$f_1$, 3$f_1$, 4\fu k.c.'s.
Bottom: Hintz \& Joner (1997) measurements, \fu, 2$f_1$, 3$f_1$, 4\fu k.c.'s.
The presence of a residual signal is quite evident in the middle and bottom
panels. \label{fig5}}
 \end{figure}
\clearpage
\begin{figure}
\epsscale{0.65}
\plotone{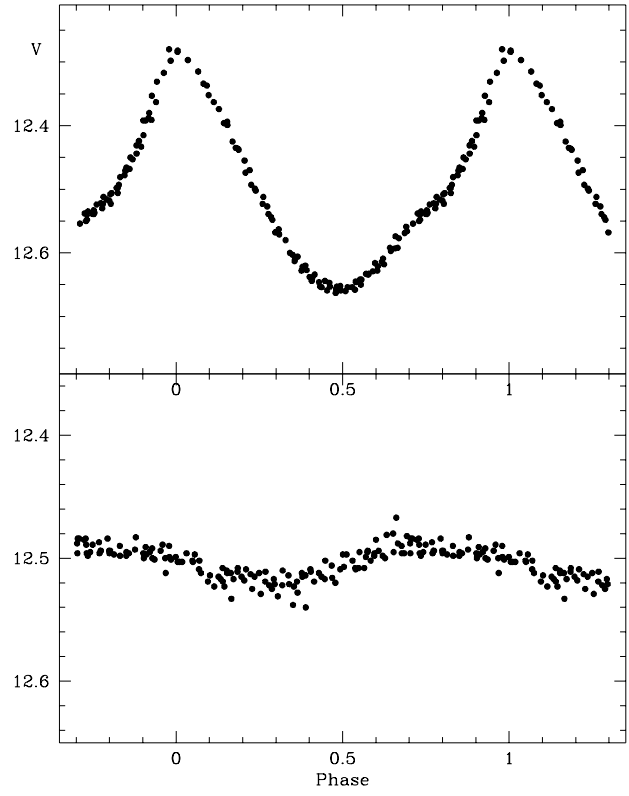}
\caption{Light curves of V798 Cyg as obtained from new CCD measurements.
Upper panel: first frequency (5.134195 \cd); lower panel: second
frequency (6.409439 \cd). \label{fig6}}
 \end{figure}

\clearpage
\begin{figure}
\epsscale{0.58}
\plotone{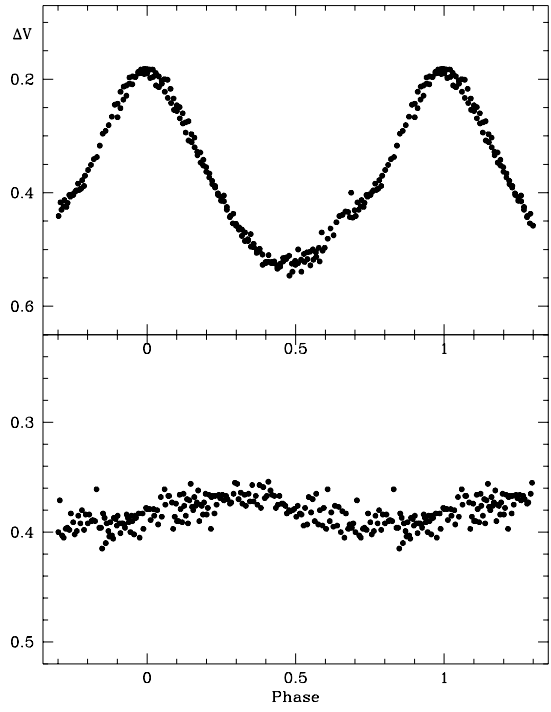}
\caption{CCD measurements of V798 Cyg by Hintz \& Joner (1997) re-interprated
in the double--mode scenario.
Upper panel: phased with $f_1$, $f_2~$
subtracted from the data. Bottom: phased with $f_2$, \fu subtracted from
the data. \label{fig7}}
 \end{figure}
\clearpage

\begin{figure}
\epsscale{0.62}
\plotone{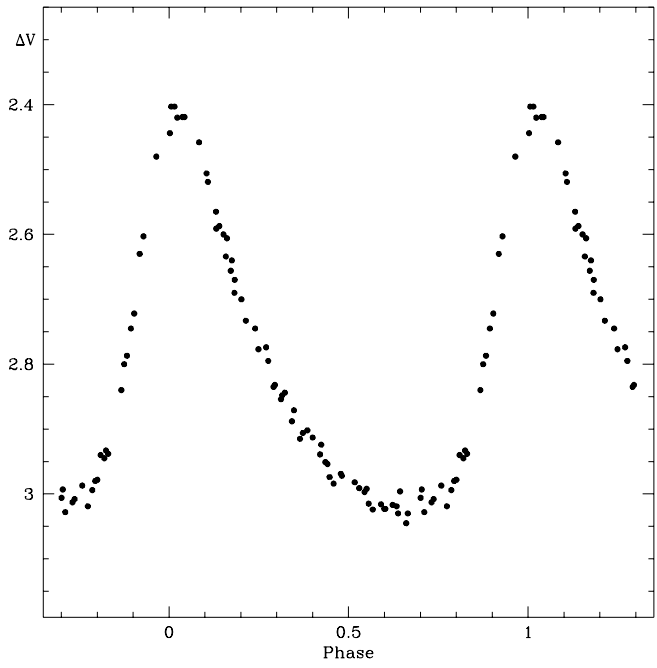}
\caption{The light curve of V831 Tau as obtained from CCD measurements
carried out at San Pedro M\'artir Observatory; the period is 0.0643 d, i.e. 92.6
minutes. \label{fig8}}
\end{figure}
\clearpage

\begin{deluxetable}{l c rr c rr c rr}
\footnotesize
%\begin{table*}
%\begin{center}
\tablecaption{Least--squares fit of the three datasets available on V798 Cyg.
The Loiano data are in an instrumental $V$ system; the Hintz \& Joner values
are magnitude differences}
\tablewidth{0pt}
%\begin{tabular}{l c rr c rr c rr}
%\tableline
%\noalign{\smallskip}
\tablehead{ & &\multicolumn{2}{c}{Photoelectric} 
& &\multicolumn{2}{c}{Loiano CCD} 
& &\multicolumn{2}{c}{Hintz \& Joner CCD} \\
\cline{3-4}
\cline{6-7}
\cline{9-10}
\multicolumn{1}{c}{Term} & &\multicolumn{1}{c}{$A_i$} & \multicolumn{1}{c}{$\phi_i$}
& &\multicolumn{1}{c}{$A_i$} & \multicolumn{1}{c}{$\phi_i$}
& &\multicolumn{1}{c}{$A_i$} & \multicolumn{1}{c}{$\phi_i$}\\
\multicolumn{1}{c}{[\cd]} & &\multicolumn{1}{c}{[mag]} & \multicolumn{1}{c}{[rad]}
& &\multicolumn{1}{c}{[mag]} & \multicolumn{1}{c}{[rad]}
& &\multicolumn{1}{c}{[mag]} & \multicolumn{1}{c}{[rad]}\\  }
%\noalign{\smallskip}
%\tableline
%\noalign{\smallskip}
\startdata
\fu = 5.134915 & & 0.182\t0.002 & 3.19\t0.01 & & 0.165\t0.001 & 3.05\t0.01 & & 0.159\t0.001 & 3.19\t0.01 \\
2\fu           & & 0.032\t0.002 & 2.62\t0.06 & & 0.032\t0.001 & 2.46\t0.03 & & 0.027\t0.001 & 2.80\t0.04 \\
3\fu           & & 0.019\t0.002 & 3.34\t0.09 & & 0.018\t0.001 & 2.99\t0.05 & & 0.012\t0.001 & 3.48\t0.09 \\
4\fu           & & 0.005\t0.002 & 4.53\t0.35 & & 0.008\t0.001 & 3.48\t0.12 & & 0.004\t0.001 & 4.73\t0.29 \\
\fd = 6.409439 & & 0.012\t0.002 & 0.51\t0.16 & & 0.016\t0.001 & 4.57\t0.06 & & 0.013\t0.001 & 1.20\t0.08 \\
%\noalign{\smallskip}
$V_0$          & & \multicolumn{2}{c}{12.504\t0.001} & &\multicolumn{2}{c}{12.504\t0.001} & & \multicolumn{2}{c}{0.3815\t0.001}\\
Residual rms   & & \multicolumn{2}{c}{0.018 mag} & &\multicolumn{2}{c}{0.007 mag} & & \multicolumn{2}{c}{0.009 mag}\\
$N$          & & \multicolumn{2}{c}{204} & &\multicolumn{2}{c}{134} & & \multicolumn{2}{c}{182}\\
$T_0$ (HJD)    & & \multicolumn{2}{c}{2446639.2526} & &\multicolumn{2}{c}{2447036.5840} & & \multicolumn{2}{c}{2449621.5182}\\
%\noalign{\smallskip}
%\tableline
\enddata
%\end{tabular}
%\end{center}
%\end{table*} 
\end{deluxetable}
\clearpage

\begin{deluxetable}{l c rr}
\footnotesize
\tablecaption{Least--squares fit of the measurements of V831 Tau}
\tablewidth{0pt}
\tablehead{
\multicolumn{1}{c}{Term} & &\multicolumn{1}{c}{$A_i$} & \multicolumn{1}{c}{$\phi_i$}\\
\multicolumn{1}{c}{[\cd]} & &\multicolumn{1}{c}{[mag]} & \multicolumn{1}{c}{[rad]}\\
}
\startdata
\fu = 15.5528 & & 0.270\t0.003 & 2.66\t0.02\\
2\fu              & & 0.098\t0.003 & 2.79\t0.05\\
3\fu              & & 0.039\t0.004 & 2.72\t0.14\\
4\fu              & & 0.014\t0.003 & 3.39\t0.39\\
$\Delta V_0$          & & \multicolumn{2}{c}{2.809\t0.002}\\
Residual rms   & & \multicolumn{2}{c}{0.014 mag}\\
$N$          & & \multicolumn{2}{c}{81}\\
$T_0$ (HJD)    & & \multicolumn{2}{c}{2450062.050}\\
\enddata
\end{deluxetable}

\end{document}